\documentclass[12pt]{article}
\usepackage{amsfonts}

\begin{document}

\title{\bf The Friedrichs-Model with fermion-boson couplings}

\author{
{O. Civitarese$^1$
\thanks{e-mail: civitare@fisica.unlp.edu.ar} and
M. Gadella $^2$. \thanks{e-mail: gadella@fta.uva.es}}\\
{\small\it$^{1}$ Departamento de F\'{\i}sica, Universidad Nacional
de La Plata, }
\\ {\small\it c.c. 67 1900, La Plata, Argentina. }
\\ {\small\it$^{2}$Departamento de F\'{\i}sica Te\'orica, Facultad de Ciencias,
E-47011, Valladolid, Spain}\\}

\maketitle

\noindent {\small {\bf Abstract}: In this work we present an
extended version of the Friedrichs Model, which includes
fermion-boson couplings. The set of fermion bound states is
coupled to a boson field with discrete and continuous components.
As a result of the coupling some of the fermion states may become
resonant states. This feature suggests the existence of a formal
link between the occurrence of Gamow Resonant States in the boson
sector, as predicted by the standard Friedrichs Model, with
similar effects in the set of solutions of the fermion central
potential (Gamow fermion resonances). The structure of the
solutions of the model is discussed by using different
approximations to the model space. Realistic couplings constants
are used to calculate  fermion resonances in a heavy mass nucleus.
}
\\


\section{Introduction}

The conventional approach to the quantum mechanical description of
the nuclear many body problem is, of course, the shell model
\cite{x1}. The task of constructing many-body nuclear wave
functions, starting from the shell model treatment of two-body
interactions, is formidable because of the extremely large
dimension of the basis, and only very recently some impressive
steps towards that goal have been achieved \cite{x2}.
Alternatively, one may think of physical oriented approximations
to the problem, based on the use of effective one-particle and
collective degrees of freedom and their couplings \cite{BM1,BM2}.
Central to this picture is the role play by one particle fermion
bound states, collective boson excitations (either vibrational or
rotational) and particle-boson couplings \cite{BM2}.

In both approaches, either the shell model or the one-particle
plus collective excitations, single-particle states are obtained
as solutions of nuclear central potentials, like the nuclear
harmonic oscillator or the Woods-Saxon potential \cite{BM1}. From
the mathematical point of view these are solutions of the
Schroedinger equation with a central potential with volume,
surface and spin-orbit interactions \cite{x3}. Since the nucleus
is a system with a finite density of one-particle states, one is
interested in one-particle bound state solutions. Shell model
basis are thus composed by these single-particle states \cite{x4}.
Since the one-particle potentials are of finite range,
single-particle resonant states are obtained by imposing purely
outgoing boundary conditions \cite{x6}.

Tore Bergreen \cite{Berg1,Berg2} was one of the first physicist to
show that a single-particle basis may contain not only bound
states but also some single-particle resonances. Lately, the work
of many others nuclear structure physicist followed
\cite{x7,x8,x9} and by now the subject is rather well established
and several nuclear structure calculations have been performed in
Bergreen's basis to describe nuclear structure properties in the
continuum, decay widths of particle states and vibrational states,
nuclear clusters and their decays \cite{kruppa}.

Since the calculations involved a rather large number of states,
even in the simplest versions, one may be tempted to study the
effects of couplings between bound and resonant states in the
context of solvable models, in order to extract signatures of
possible dominant mechanisms.

In the last 40 years, the use of solvable models has contributed
significantly to the understanding of the nuclear dynamics, in
spite of its complexity. Although solvable models are popular in
the context of bound single-particle and collective states
\cite{RS,x10}, their use to describe collective and
single-particle properties in the continuum is less frequently
found in the literature.

The Friedrichs model  \cite{1} is a suitable mathematical model to
describe the coupling with the continuum. The model has the
advantage of being exactly solvable and it can be applied to
various systems of physical interest \cite{2,3,4,5}. In its
simplest version the model deals with the coupling between
discrete and continuous components of a boson field \cite{1}. More
sophisticated versions of the Friedrichs model have been presented
in \cite{2,3,6}. It is the only completely solvable model which
describes resonance phenomena \cite{7}. A recent presentation of
the model, in connection with the study of resonances in field
theory, can be found in \cite{4}. In this work we aim at the
extension of the model, in order to include the coupling between
fermion and boson fields.

The advantage of using an exactly solvable model, like the
Friedrichs model, to study the coupling with the continuum is
rather obvious, since until now the description of these couplings
are limited to purely numerical approaches. In the standard
approach, like the one advocated in the literature
\cite{x7,x10,x9}, the occurrence of resonances in the fermion
sector is a result of the boundary conditions imposed to the
diagonalization of finite range central potentials. Purely
outgoing boundary conditions lead to Gamow resonances; some of
these states, depending upon the imaginary part of their energies,
may be included in single-particle basis, the so-called Berggren
basis, together with fermion bound-states. In this hibrid basis
one may attempt to include residual two-body interactions. As
shown \cite{x9} in the energies of some particle-hole excitations
may become complex. The same effect shows up at the level of
attractive two-particle configurations \cite{rolo,hm}, where
resonances may appear as a consequence of the coupling between
configurations of fermion pairs in bound and resonant states.
Finally, it is worth mentioning that the observed properties of
atomic nuclei in the drip line may be strongly linked to the
coupling between bound and resonant states. Particularly, we refer
to the broadening of the single particle energies and to the
appearance of a decay width in single-particle orbits, which may
be strongly influenced by the coupling to continuous particle-hole
excitations \cite{hm,cdl}. A possible mechanism to explain for
this class of phenomena is the coupling between bound fermion
states and discrete and continuous bosonic degrees of freedom.

Since the standard Friedrichs model describes the coupling between
the discrete and the continuous components of a single boson
field, it seems natural to include fermions and their couplings
with the boson field and take advantage of the solvable structure
of the model. In this respect, the results of the extended
Friedrichs model may be used to approximate some specific features
of fermion states, like their decay width, in situations where the
coupling between fermions and bosons indicates the existence of
effects due to the continuum \cite{cdl}.

 The present paper is organized as follows: In Section 2, we
present the needed mathematical details about the model. In
Section 3, we explore the structure of the solution obtained under
different approximations. We have applied a compact formalism to
solve the system of coupled equations yielded by the model. The
results of applications, with reference to a realistic case, are
presented and discussed in Section 3. Conclusions are drawn in
Section 4.

\section{The Model}\label{s1}

For the benefit of the reader, we shall begin with a review of the
basic ideas of the Friedrichs Model, which is an exactly solvable
model where a discrete boson state interacts with a continuum
spectrum of boson excitations. The situation resembles very much
that of the nuclear excitations of even mass nuclei near the drip
line, where few bound states may strongly interact with low-lying
continuum states. This model provides the insight about the
mechanism leading to the appearance of resonances in the boson
sector, and it has the clear benefit of being solvable.

Next, we shall review briefly the properties of resonant solutions
of the Friedrichs model and then we shall proceed with the
description of the extension which we propose in this paper. It
consists of the inclusion of fermion bound states which interact
with the boson resonant state. Since the extended model is also
solvable, one might also get some insight about the effects of the
coupling of boson and fermion states in the vicinity of the nucleon
drip line. From the mathematical point of view the extension of the
Friedrichs model amounts to the simultaneous treatment of fermion
and boson sectors of a Hamiltonian which includes resonant states in
its spectrum. From the point of view of physics it represents the
solvable limit of nucleon-boson interactions in  Berggreen's basis
\cite{cdl}.

In spite of some formal steps of the derivation, the present
discussion is motivated by the need to build a schematic, albeit
solvable, model which may be use to approach dominant features of
the particle-vibration coupling in the continuum.

Further details about the formalism concerning the standard
Friedrichs model can be found in \cite{AGPP, AGP} and references
quoted therein.

\subsection{The Standard Friedrichs Model}\label{ss1}

The simplest form of the Friedrichs model \cite{1} includes a free
Hamiltonian $H_0$ with a real positive continuous spectrum ($\omega
> 0$) and a discrete eigenvalue ($\omega_0 >0$) imbedded in the
continuous spectrum. An interaction is acting between the continuous
and discrete parts of $H_0$ by means of a potential $V$. As a result
of the action of $V$, the bound state of $H_0$ is dissolved in the
continuous and a resonance is produced. The spectrum of the total
Hamiltonian $H=H_0+ \lambda V$, where $\lambda$ is a real coupling
constant, is purely continuous and coincides with the real positive
semiaxis. We write for the free Hamiltonian, $H_0$, the expression

\begin{equation} H_0=\omega_0|1\rangle\langle 1|+\int_0^\infty
\omega |\omega\rangle\langle\omega|\,d\omega\,. \label{1}
\end{equation}
The interaction, $V$, is given by

\begin{equation}
V=\int_0^\infty [V^*(\omega)\,|
\omega\rangle\langle1|+V(\omega)\,|1\rangle\langle\omega|]\,d\omega\,,
\label{2}
\end{equation}
where, $V(\omega)$ is called the form factor \cite{3,HMR}.

The most general form of the state vector $\psi$, in the basis
spanned by $\{|1\rangle, |\omega\rangle\}$, can be written as

\begin{equation} \psi=\alpha |1\rangle+\int_0^\infty
\varphi(\omega)\,|\omega\rangle\,d\omega .  \label{3}
\end{equation}
Note that the set $\{|1\rangle,|\omega\rangle\}$,
$\omega\in[0,\infty)$, of generalized eigenvectors of $H_0$, is a
complete set. Thus, each vector state $\psi$ has a component on the
bound state $|1\rangle$ of $H_0$ and an infinite number of
components, in the continuous part of the spectrum of $H_0$. The
action of $V$ on $\psi$ is given by

\begin{equation}
V\psi=\left( \int_0^\infty d\omega
V(\omega)\varphi(\omega)\right)|1\rangle+ \alpha \int_0^\infty
d\omega V^*(\omega)|\omega \rangle\,.
\end{equation}

To obtain the explicit form of (\ref{3}), that is to determine the
amplitudes $\alpha$ and $\varphi(\omega)$, we shall use the
technique presented in \cite{AGP}\footnote{See also \cite{5},
section 3.5.}. In the notation of \cite{AGP}, the reduced
resolvent, $\eta(z)$ is given by (see also \cite{1,HMR})

\begin{equation}
\eta(z)=\langle 1|\frac{1}{H-z}|1\rangle\,.
\end{equation}
By making adequate assumptions on the form-factor $V(\omega)$,
$\eta(z)$ has the following properties \cite{5}:

i.) The function $\eta(z)$ has no singularities in the complex
plane other than a branch cut coinciding with the positive
semi-axis, i.e., the continuous spectrum of $H$.

ii.) The function $\eta(z)$ can be analytically continued through
the cut. These are the functions $\eta_{\pm}(z)$

\begin{equation}
\eta_{\pm}(z)=\langle 1|\frac{1}{H-\omega\pm i0}|1\rangle
=\omega_0 -z-\int_0^\infty
\frac{\lambda\,|V(\omega)|^2\,d\omega}{z-\omega\pm i0}\,.
\label{6}
\end{equation}
These extensions have poles located at the points
$z_0=E_R-i\frac{\Gamma}{2} $, for $\eta_-(z)$, and $z_0^*$, for
$\eta_+(z)$, with $E_R >0$ and $\Gamma >0$.

Although resonance poles may, in principle, have arbitrary
multiplicity  \cite{M}, depending on $V(\omega)$, we shall assume
hereafter that the resonance poles are simple.

\subsection{Resonance or Gamow states.}

Poles of (\ref{6}) appear into complex conjugate pairs, each of them
represents a resonance. Then, Gamow vectors are solutions of the
eigenvalue equation,

\begin{equation}
(H-x)\,\Psi(x)=0\,. \label{150}
\end{equation}
with the total Hamiltonian $H$, whose eigenvalues coincide with
resonance poles \cite{AGP}.

Since $|1\rangle$ and $|\omega\rangle$ form a complete system, we
must have:

\begin{equation}
\Psi(x)= \alpha(x)\,|1\rangle + \int_0^\infty
\psi(x,\omega)\,|\omega\rangle\,d\omega\,. \label{151}
\end{equation}
If we apply $H$ to (\ref{151}), we obtain the following system of
equations:

\begin{equation} (\omega_0-x)\alpha(\omega)+\lambda
\int_0^\infty \psi(x,\omega)\,V^*(\omega)\,d\omega =0
\,,\label{152}
\end{equation}

\begin{equation}
(\omega-x)\,\psi(x,\omega)+\lambda V(\omega)\,\alpha(\omega)=0
\,.\label{153}
\end{equation}
To solve this system, we write $\alpha(\omega)$ in terms of
$\psi(x,\omega)$ using (\ref{153}) and carry the result to
(\ref{152}). We obtain an integral equation which gives one solution
of the form:

\begin{equation} \Psi_+(x)= |x \rangle+\lambda
V^*(x)\,\frac{1}{\eta_+(x)}\left\{ |1 \rangle + \lambda
\int_0^\infty \frac{V(\omega)}{x-\omega+i0}\,|\omega \rangle
\,d\omega \right\}\,. \label{154} \end{equation} The bracket $
\Psi_+(x)$, with an arbitrary vector of the form $\psi$, gives a
function which is analytically continuable from above to below
through the spectrum of the Hamiltonian $H$. This continuation has
a simple pole at $z_0$ so that we can write on a neighborhood of
$z_0$:

\begin{equation} \Psi_+(z)= \frac{C}{z-z_0}+o(z)\,. \label{155}
\end{equation}
 From (\ref{150}) and (\ref{155}), we get

\begin{equation}
(H-z)\Psi_+(z)=\frac{1}{z-z_0}\; (H-z) \,C+(H-z)\,o(z)=0 \,,
\label{156} \end{equation} which gives

\begin{equation}
(H-z_0)\,C=0 \Longrightarrow HC=z_0 C\,. \label{157}
\end{equation}
This shows that the residue $C$ of $\Psi_+(z)$ at the pole $z_0$
coincides, save for an irrelevant constant, with the decaying
Gamow vector $|f_0\rangle$\footnote{Each resonance is
characterized by a pair of complex conjugate poles,
$z_0=E_R-i\Gamma/2$ and $z_0^*=E_R+i\Gamma /2$, in the
continuation of $\eta(z)$ from above to below, $\eta_-(z)$,  and
from below to above, $\eta_+(z)$, respectively. Both are
eigenvalues of $H$ and the corresponding eigenvectors are the
decaying Gamow vector, $|f_0\rangle$, with $z_0$, and the growing
Gamow vector, $|\widetilde f_0\rangle$, with $z_0^*$.}. To
calculate its explicit form note that (\ref{155}), on a
neighborhood of $z_0$, has the form:

\begin{equation} \Psi_+(z)\approx \frac{\rm {constant}}{(z-z_0)} \;\left\{
|1\rangle+ \lambda \int_0^\infty
\frac{V(\omega)}{z-\omega+i0}\,|\omega\rangle\,d\omega \right\}
+{\rm RT}\,, \label{158} \end{equation} where RT stand for
``regular terms''.  Now, let us use the Taylor theorem to write:

\begin{equation} \frac{1}{z-\omega+i0} = \frac{1}{z_0-\omega+i0}
- \frac{z-z_0}{(z_0-\omega+i0)^2} +o(z)\,. \label{159}
\end{equation} By replacing (\ref{159}) in (\ref{158}), we get

\begin{equation} \Psi_+(z)\approx \frac{\rm {constant}}{(z-z_0)} \;
\left\{ |1\rangle+ \lambda \int_0^\infty
\frac{V(\omega)}{z_0-\omega+i0}\,|\omega\rangle\,d\omega \right\}
+{\rm RT}\,. \label{160} \end{equation} Therefore, up to an
irrelevant constant, we conclude that

\begin{equation} C\equiv |f_0\rangle=|1\rangle+ \int_0^\infty
\frac{\lambda
\,V(\omega)}{z_0-\omega+i0}\,|\omega\rangle\,d\omega\,.
\label{161}
\end{equation} The system given by equations (\ref{152}) and
(\ref{153}) has another solution that can be analytically continued
in the upper half plane. This solution gives, using the same
technique, the growing Gamow vector $|\widetilde f_0\rangle$:

\begin{equation} |\widetilde f_0\rangle= |1\rangle+ \int_0^\infty
\frac{\lambda
\,V^*(\omega)}{z^*_0-\omega-i0}\,|\omega\rangle\,d\omega\,.
\label{162}
\end{equation}

So far, we have introduced the simplest and more standard version
of the Friedrichs model in a fashion accessible to the reader.
This will facilitate the comprehension of the more complicated
version under consideration in the next subsection.

\subsection{The extended Friedrichs Model.}\label{ss21}

The present version of the Friedrichs model, is an extension of
the previously introduced standard Friedrichs model of Section 1.2
and it includes :

i) The unperturbed fermion and boson Hamiltonian $H_I$

\begin{eqnarray}
H_{I}=\omega_0|1\rangle \langle 1| + \int_0^\infty d\omega \omega
|\omega \rangle \langle \omega | +\sum_k c_k |k\rangle \langle
k|\,,\label{20}
\end{eqnarray}
where the index $k$ runs out the set of Fermion kets $|k\rangle$.

ii) The interaction between fermions, $|k\rangle$, and the discrete
boson, $|1\rangle$, Hamiltonian $H_{II}$:

\begin{equation}
H_{II}=\sum_{k,l}\left[h_{k,l}|k,1\rangle \langle l|+
h^*_{k,l}|l\rangle \langle k,1|\right]\,.\label{21}
\end{equation}

iii) The interaction between fermions and the boson field,
$\{|\omega\rangle\}$, $H_{III}$:

\begin{equation}
H_{III}=\sum_{k,l}\int_0^\infty d\omega \left[f_{k,l}(\omega)
|k,\omega \rangle\langle l|+ f^*_{k,l}(\omega) |l\rangle\langle
k,\omega|\right]\,.\label{22}
\end{equation}

 The standard Friedrichs model includes the boson-boson coupling $V$ described by
equation (\ref{2}). This coupling $V$ can be generalized to
include fermion-boson interactions in the following manner
\cite{BM2,W}

\begin{equation}
H_{IV}=\sum_{k,k'}\int_0^\infty d\omega \left[g_{kk'}(\omega) |k,1
\rangle\langle k',\omega|+ g^*_{kk'}(\omega)
|k',\omega\rangle\langle k,1|\right]\,.\label{23}
\end{equation}
To show that $H_{IV}$ generalizes $V$, of Eq.(2), it is enough to
replace the coupling $g_{kk'}(\omega)$ by $g(\omega)$ for all values
of $k$ and use the identity $\sum_k | k \rangle \langle k|=1$.

In the following, we shall use $g(\omega)=V(\omega)$, for which
$H_{IV}$ coincides with the standard Friedrichs model interaction
$V$.

To obtain the solution of the eigenvalue problem

\begin{equation}
(H-E)\Psi(E)=0\,,
\end{equation}
where $H= H_I+H_{II}+H_{III}+V$, we write $\Psi(E)$ in its most
general form, as

\begin{equation}
\Psi(E)=\sum_k\varphi_k(E)|k\rangle+\sum_k \phi_{k,1}(E)|k,1\rangle
+\sum_k\int_0^\infty d\omega \,\psi_k(E,\omega)|k,\omega\rangle\,.
\end{equation}
The action of the Hamiltonian on $\Psi(E)$ yields

\begin{eqnarray}
&&(H-E)\Psi(E)\equiv \sum_k\varphi_k(E)(c_k-E)|l\rangle
+\sum_{k,l} h_{kl}^*\phi_{k,1}(E)|k\rangle \nonumber \\
&+&\sum_k (c_k+\omega_0-E)\phi_{k,1}(E)|k,1\rangle
+\sum_k\int_0^\infty d\omega \psi_k(E,\omega)V(\omega)|k,1\rangle
\nonumber
\\ &+&\sum_k \phi_{k,1}(E)\int_0^\infty d\omega
V^*(\omega)|k,\omega\rangle +\sum_k\int_0^\infty d\omega
\psi_k(E,\omega)(c_k+\omega-E)|k,\omega\rangle\nonumber
\\ &+&\sum_{k,l}\int_0^\infty d\omega
\psi_k(E,\omega)f_{kl}^*(\omega)|l\rangle
+\sum_{i,k}\varphi_k(E)h_{ik}|i,1\rangle\nonumber \\
&+&\sum_{i,k}\int_0^\infty d\omega
\varphi_k(E)f_{ik}(\omega)|i,\omega\rangle=0
\end{eqnarray}
The above equation can be re-arranged as a linear combination of
$|k\rangle$, $|k,1\rangle$, and $|k,\omega\rangle$ with vanishing
coefficients. Thus,

\begin{equation}
\varphi_k(E)(c_k-E)+\sum_lh_{lk}^*\phi_{l1}(E)+\sum_l\int_0^\infty
d\omega \psi_l(E,\omega)f_{lk}^*(\omega)=0 \label{27}
\end{equation}

\begin{equation}
\sum_l\varphi_l(E)h_{kl}+(c_k+\omega_0-E)\phi_{k1}(E)+\int_0^\infty
d \omega \psi_k(E,\omega)V(\omega)=0 \label{28}
\end{equation}

\begin{equation}
\psi_k(E,\omega)(c_k+\omega-E)+\sum_l\varphi_l(E)f_{kl}
(\omega)+\phi_{k1}(E)V^*(\omega)=0 \label{29}
\end{equation}
for $|k\rangle$, $|k,1\rangle$, and $|k,\omega\rangle$,
respectively. A new set of equations, where only the amplitudes
$\varphi_l(E)$ and $\phi_{l1}(E)$ appear, can be obtained from the
above system. From the last equation we obtain:

\begin{equation}
\psi_k(E,\omega)=c\delta(c_k-\omega-E)- \sum_l\frac{ \varphi_l(E)
f_{kl}(\omega)}{c_k-\omega-E}-\frac{\phi_{k1}(E)V^*(\omega)}{c_k-\omega-E}\,,
\label{30}
\end{equation}
where $c$ is an arbitrary constant.

If we replace (\ref{30}) into (\ref{27}) and (\ref{28}), we obtain
an infinite system of coupled equations in the amplitudes
$\phi_{l1}(E)$ and $\varphi_l(E)$. In the most general case, to find
the analytic solution of this system may be rather difficult.
 Note that the
resonance behavior is attained by the coupling to the bosons.  In
order to obtain solutions for the above system, we need to make
appropriate choices of the model input represented by the
couplings: $h_{kl}$, $f_{kl}(\omega)$ and $V(\omega)$.

After this substitution, we obtain the following result:

\begin{eqnarray}
\left[(c_k-E)\delta_{km}-\sum_m
A_{km}(E)\right]\varphi_m(E)\nonumber\\+\sum_m\left( h_{mk}^*-
B_{km}(E)\right)\phi_{m1}(E)=-c\sum_m f_{mk}^*(E-c_m)\label{trece}
\end{eqnarray}
and
\begin{eqnarray}
&&\sum_l [h_{kl}-\widetilde B_{kl}(E)]\varphi_l(E)\nonumber\\[2ex]&&+
(c_k+\omega_0-E-C_k(E))\phi_{k1}(E)=-cV(E-c_k)\,,\label{quince}
\end{eqnarray}
where we have introduced the following notations:

\begin{eqnarray}
A_{km}(E)=\int_0^\infty d\omega\;
\sum_l\frac{f^*_{lk}(\omega)\,f_{lm}(\omega)}{c_l+\omega-E}\label{A}\\[2ex]
B_{km}(E)=\int_0^\infty d\omega\;
\frac{f^*_{mk}(\omega)\,V^*(\omega)}{c_m+\omega-E}\label{B}\\[2ex]
\widetilde B_{km}(E)= \int_0^\infty d\omega
\;\frac{V(\omega)\,f_{km}(\omega)}{c_k+\omega-E}\label{Btilde}\\[2ex]
C_k(E)= \int_0^\infty d\omega\;
\frac{|V(\omega)|^2}{c_k+\omega-E}\label{C}
\end{eqnarray}
Notice that expressions (\ref{B}-\ref{C}) contain both $f(\omega)$
and $V(\omega)$, as a result of coupling of fermions and bosons in
the continuous sector of the model.  The solution already found in
Sections 2.1 and 2.2 is recovered if $f_{km}(\omega)=h_{km}=0$,
since $C_k(E), \neq 0 $ and then Eq.(\ref{quince}) reduces to
Eq.(\ref{6}).


\section{A simplified version of the Model: One fermion level.}

In this section, we are going to discuss a simplified version of
the Friedrichs model including the following ingredients:

i.) A single fermionic bound level. This means that there is only
one term in the sum that appears in $H_I$ in (\ref{20}). This term
is denoted as

\begin{equation}
H_{F}=\varepsilon\,|i\rangle\langle i|\,,
\end{equation}
where we write $|i\rangle$ to denote the fermion level and thus
avoid confusion with the boson bound state denoted as $|1\rangle$.

ii.) A standard Friedrichs model as described in the previous
section.

iii.) An interaction between the fermion level and the boson field
given by the following reduced version of $H_{III}$:

\begin{equation}
H_{III}= \int d\omega\, [f(\omega)\,|i\rangle\langle
i,\omega|+f^*(\omega)\,|i,\omega\rangle\langle i|\,]\,.\label{s1}
\end{equation}

Thus, the total Hamiltonian is given by

\begin{eqnarray}
H=\omega_0\,|1\rangle\langle 1|+\int_0^\infty
|\omega\rangle\langle
\omega|\,d\omega +\varepsilon\,|i\rangle\langle i|\nonumber\\[2ex] +
\lambda\int_0^\infty [V(\omega)|\omega\rangle\langle
1|+V^*(\omega)|1\rangle\langle\omega|\,]\,d\omega\nonumber\\[2ex] +\int
d\omega\, [f(\omega)\,|i\rangle\langle
i,\omega|+f^*(\omega)\,|i,\omega\rangle\langle i|\,]\,.
\end{eqnarray}

The interaction produced by $H$ is depicted in Figure 1.

Take now the part of the Hamiltonian that corresponds to the
standard Friedrichs model, given by $H_0+\lambda V$ as in
(\ref{1}) and (\ref{2}). It has been shown (\cite{G}) that this
Hamiltonian can be diagonalized in terms of the Gamow vectors and
the background part as

\begin{equation}
H=z_R\,|f_0\rangle\langle\widetilde f_0|+\int_\Gamma
z|z\rangle\langle z|\,d\mu(z)\label{s4}
\end{equation}

In the description given by (\ref{s4}), the boson bound state has
been replaced by the Gamow vector $|f_0\rangle$ and the boson
field by the background. We can ignore or neglect the background
term and then consider a two level system: one level is given by
the fermion bound state $|i\rangle$ and the other by the Gamow
vector $|f_0\rangle$. The resulting model can be seen as depicted
in Figure 2.

The situation now as visualized in Figure 2 contains two systems
in interaction for which the Hamiltonian can be written
as\footnote{As a matter of fact, this Hamiltonian should be
written as
$$H=\varepsilon  |i\rangle\langle i|\otimes I+z_R\,I\otimes
|f_0\rangle\langle\widetilde f_0| +
 \Lambda(z_R)[|i,f_0\rangle \langle i|+|i\rangle\langle
i,\widetilde f_0|\,]\,, $$ where $I$ is the identity operator. In
terms of creation and annihilation operators, this Hamiltonian can
be written as $$ H=\varepsilon c_i^\dagger c_i+z_RG^\dagger
G+\Lambda(z_R)[c_i^\dagger G^\dagger c_i+c_i^\dagger Gc_i]\,,
$$ where $c_i^\dagger$ and $c_i$ are the creation and annihilation
operators for the fermion with state $|i\rangle$ and
$G^\dagger$ and $G$ are the creation and annihilation operators
for the Gamow state $|f_0\rangle$.}

\begin{equation}
H=\varepsilon |i\rangle\langle i|+z_R|f_0\rangle\langle \widetilde
f_0|+\Lambda(z_R)[|i,f_0\rangle \langle i|+|i\rangle\langle
i,\widetilde f_0|\,]\,,\label{s5}
\end{equation}
where the symbol $\Lambda(z_R)$ denotes a complex number which
depends on the position of the resonance (and therefore on the
value of the coupling constant $\lambda$) only.

It is important to remark that in (\ref{s5}) we have neglected the
contribution of the background integral, (a term of the form
$\int_\Gamma z|z\rangle\langle \widetilde z|\,d\mu(z)$) and note
that consequently the resulting Hamiltonian is not formally
self-adjoint.

Our goal is to show the existence of resonances, i.e., complex
solutions in $\eta$ of the eigenvalue equation

$$
H\varphi=\eta\varphi\,.
$$

The most general form of the vector $\varphi$ can be given as:

\begin{equation}
\varphi=\alpha|i,0\rangle+\beta|i,f_0\rangle\,,
\end{equation}
where: i.) The vector $|i,f_0\rangle$ is the tensor product
$|i\rangle\otimes|f_0\rangle$. ii.) The vector $|i,0\rangle$ is
the tensor product $|i\rangle\otimes|0\rangle$, where $|0\rangle$
denotes a state with zero resonances. In the basis given by
$\{|i,0\rangle,\,|i,f_0\rangle\}$ the matrix elements of $H$ can
be readily obtained:

\begin{eqnarray}
H=\left(\begin{array}{cc} \langle i,0|H|i,0\rangle & \langle
i,0|H|i,f_0\rangle\\[2ex] \langle i,\widetilde
f_0|H|i,0\rangle & \langle i,\widetilde
f_0|H|i,f_0\rangle\end{array}\right)= \left(\begin{array}{cc}
\varepsilon & \Lambda(z_R)\\[2ex] \Lambda(z_R) &
\varepsilon+z_R\end{array}\right)\,.\label{s7}
\end{eqnarray}

In order to obtain the eigenvalues of $H$, we need to solve the
equation $\det(H-E)=0$ and this gives:

\begin{equation}
\det(H-E)=(\varepsilon-E)(\varepsilon+z_R-E)-\Lambda^2(z_R)=0\,,
\end{equation}
i.e.,

\begin{equation}
E^2-(2\epsilon+z_R)E+z_R\epsilon-\Lambda^2(z_R)=0\,,
\end{equation}
which gives

\begin{equation}
E=\varepsilon+\frac{z_R}{2}\pm\frac12
\sqrt{z_R^2+\Lambda^2(z_R)}\,.
\end{equation}
Observe that, for any given resonance $z_R$ in the boson levels,
there exists two new resonances in the coupled boson-fermion
system.

\subsection{Calculation of $\Lambda(z_R)$.}

In the simplified version under our consideration, only the
interaction Hamiltonians $H_{III}$ and $H_{IV}$ remain. For the
former, we keep only one term in the sum in (\ref{22}) so that it
has now the form:

\begin{equation}
H_{III}=\int_0^\infty [f(\omega)|i,0\rangle\langle
i,\omega|+f^*(\omega)|i,\omega\rangle\langle
i,0|\,]\,d\omega\,.\label{s11}
\end{equation}
Concerning $H_{VI}$, it only survives a term of the sum in
(\ref{23}) and the corresponding form factor (the $g(\omega)$) is
the form factor of the boson-boson field coupling $V(\omega)$. As it
was previously clarified, this coupling is the responsable for the
creation of the boson resonance given by $|f_0\rangle$.

We now want to determine the number\footnote{We write it as a
function of the position of the resonance pole, although we
equally can write it as a function of the value of the coupling
constant $\lambda$ which determines the pole, i.e., as
$\Lambda(\lambda)$.} $\Lambda(z_R)$ in the approximation that
neglects the background. Note that (\ref{s5}) is, save for the
background term equal to $H_I +H_{IV}+H_{III}$, with only one
fermion bound term in the fermion sum in $H_I$. Since
$\varepsilon|i\rangle\langle i|+z_R|f_0\rangle \langle\widetilde
f_0|+{\rm background}= H_I+H_{IV}$, we have that

\begin{equation}
H_{III}=\Lambda(z_R)[|i,f_0\rangle \langle i,0|+|i,0\rangle\langle
i,\widetilde f_0|\,]\,. \label{s12}
\end{equation}
If we multiply (\ref{s12}) to the right by the ket
$|i,f_0\rangle$, and take into account that $\langle
i,0|i,f_0\rangle=0$ and $\langle i,\widetilde
f_0|i,f_0\rangle=\langle i|i\rangle\langle \widetilde
f_0|f_0\rangle=1$ \cite{G,CGB}, we obtain:

\begin{equation}
\int f(\omega) |i\rangle\langle
i,\omega|i,f_0\rangle\,d\omega=\left[\int_0^\infty
f(\omega)\langle\omega|f_0\rangle\,d\omega\right]\,|i\rangle=[\Lambda(z_R)]\,|i\rangle\,,
\end{equation}
which obviously gives:

\begin{equation}
\Lambda(z_R)= \int_0^\infty
f(\omega)\langle\omega|f_0\rangle\,d\omega\,.
\end{equation}

The value for the bracket $\langle\omega|f_0\rangle$ is readily
obtained from (\ref{161}). This gives:

\begin{eqnarray}
\Lambda(z_R)=\lambda\int_0^\infty
\frac{f(\omega)\,V(\omega)}{z_R-\omega+i0}\,d\omega\nonumber\\[2ex]=\lambda \,{\rm
PV}\int_0^\infty \frac{f(\omega)\,V(\omega)}{z_R-\omega}\,d\omega
-i\pi\lambda f(z_R)\,V(z_R)\,.\label{s15}
\end{eqnarray}
Here, PV stands for principal value (although since $z_R$ is
complex the principal value is here the value of the integral). We
are always assuming that both form factors, $f(\omega)$ and
$V(\omega)$, are analytically continuable and that have a certain
well defined value at $z_R$.

Neglecting the integral term in (\ref{s15}) is an approximation
similar to neglecting the background term. From this point of
view, we arrive at

\begin{equation}
\Lambda(z_R)\approx -i\pi\lambda f(z_R)V(z_R)\,. \label{s16}
\end{equation}
Note that, after (\ref{s16}), $\Lambda(z_R)$ is complex.

\subsection{The effect of multiple fermion levels.}

We are going to introduce a generalization to the above discussed
simplified model. Since we are discussing decay in nuclei, it
seems natural to add multiple levels for the fermion bound states.
Now, in the extended Friedrichs model we keep the following terms:

1.- The free Hamiltonian $H_I$ as in (\ref{20}).

2.- The interaction between fermions and the boson $H_{II}$ as in
(\ref{21}).

3.- The interaction between fermions and the boson field $H_{III}$
given by (\ref{22}).

4.- The interaction boson, boson field given by (\ref{2}). Again,
note that this is a simplified form of $H_{IV}$.

The strategy is the same as in the previous subsection: we replace
the boson-boson field by the resonance-background and we neglect
the background field keeping the Gamow state. Then, we have an
interaction between the bound state fermions and the boson
resonance given by\footnote{If $c_k^\dagger$ and $c_k$
respectively are the creation and annihilation operators for the
$k-th$ fermion and $G^\dagger$ and $G$ the creation and
annihilation of the Gamow, then, this interaction Hamilonian can
be written in terms of these operators as: $$ H_{\rm
int}=\sum_{kl}\Lambda_{kl}c_k^\dagger c_k(G^\dagger+G)\,.
$$ We can operate either in the language of vectors, as shall do
in the main text or in the language of creation and annihilation
operators as is usual in the second quantization formalism.}

\begin{equation}
H_{\rm int}=\sum_{kl}\Lambda_{kl}(z_R)[|k,f_0\rangle\langle
l,0|+|k,0\rangle\langle k,\widetilde f_0|\,]\,.\label{s17}
\end{equation}

Repeating the procedure of the previous subsection, we identify
the interaction Hamiltonian, excluding $\lambda V$, with $H_{\rm
int}$ given in (\ref{s17}). This gives the following identity:

\begin{eqnarray}
H_{\rm int}=\sum_{kl}[h_{kl}|l,0\rangle\langle
k,1|+h_{kl}^*|k,1\rangle\langle l,0|\,]\nonumber\\[2ex]
+\sum_{kl} \int_0^\infty [f_{kl}(\omega)|l,0\rangle\langle
k,\omega|+f_{kl}^*(\omega)|k,\omega\rangle\langle
l,0|\,]\,.\label{s18}
\end{eqnarray}
Multiplying (\ref{s18}) to the right by $|m,f_0\rangle$, we
obtain:

\begin{eqnarray}
\sum_{kl}h_{kl}|l,0\rangle\langle k,1|m,f_0\rangle+
\sum_{kl}\int_0^\infty f_{kl}(\omega)|l,0\rangle\langle
k,\omega|m,f_0\rangle\,d\omega\nonumber\\[2ex]
=\sum_{kl}\Lambda_{kl}|l,0\rangle\langle k,\widetilde
f_0|m,f_0\rangle\,.\label{s19}
\end{eqnarray}

Since $\langle 1|f_0\rangle=1$ and $\langle
k|m\rangle=\delta_{km}$, (\ref{s19}) yields to

\begin{eqnarray}
\sum_l\{h_{ml}+\int_0^\infty
f_{ml}(\omega)\langle\omega|f_0\rangle\,d\omega\}|l,0\rangle
=\sum_l\Lambda_{ml}(z_0)|l,0\rangle\,.\label{s20}
\end{eqnarray}
As the $|l,0\rangle$ are linearly independent, (\ref{s20}) gives
for all $m,l$:

\begin{eqnarray}
\Lambda_{ml}(z_0)=h_{ml}+\int_0^\infty
f_{ml}(\omega)\langle\omega|f_0\rangle\,d\omega\nonumber\\[2ex]
\approx h_{ml}-i\pi f_{ml}(z_0)\lambda V(z_0)\,,
\end{eqnarray}
where again we have used (\ref{161}) for the determination of the
value of $\langle\omega|f_0\rangle$ and have neglected the term
principal value as in (\ref{s16}).

The space of pure states is formed by the vectors of the
form\footnote{It we use the second quantization language, the
vector can be written as $\varphi(E)=\sum_k\alpha_k
c_k^\dagger|0\rangle+\sum\beta_k c_k^\dagger G^\dagger |0\rangle$,
where $|0\rangle$ represents the state of the vacuum.}

\begin{equation}
\varphi(E)=\sum_k\alpha_k|k,0\rangle+\sum\beta_k|k,f_0\rangle\,.\label{s22}
\end{equation}

Then, we want to find the explicit form of the Hamiltonian in
terms of the basis given by the $|k,0\rangle$ and the
$|k,f_0\rangle$ and then obtaining the eigenvalues of this
Hamiltonian. The presence of complex eigenvalues of this
Hamiltonian will be equivalent to the presence of resonances for
the model.

In order to show in detail the calculation, we split the
Hamiltonian into three terms as follows:

\begin{eqnarray}
H_1=\sum_k \varepsilon |k,0\rangle\langle k,0|\,,\label{s23}\\[2ex]
H_2=z_R|0,f_0\rangle\langle 0,\widetilde f_0| \,,\label{s24}\\[2ex]
H_3= \sum_{kl} \Lambda_{kl}|k,f_0\rangle\langle l,\widetilde f_0|
+ H_{\rm BG}\,,
\end{eqnarray}
where $H_{\rm BG}$ denotes the background part of the Hamiltonian
that we neglect and we have written $\Lambda_{kl}$ instead of
$\Lambda_{kl}(z_0)$ in order to simplify the notation. After
simple manipulations, we obtain the following\footnote{We arrive
to the same results using second quantization notation. In this
notation, $H_1=\sum_k\varepsilon_kc_k^\dagger c_k$,
$H_2=z_RG^\dagger G$ and
$H_3=\sum_{kl}\Lambda_{kl}(G^\dagger+G)c_k^\dagger c_l$. We use
the anti-commutation relation $\{c_k,c_l^\dagger\}=\delta_{kl}$
and the commutation relation $[G,G^\dagger]=1$ and note that the
action of $C^\dagger$ and $G^\dagger$ in the vacuum $|0\rangle$ is
given by $c^\dagger|0\rangle=|k,0\rangle$ and
$G^\dagger|0\rangle=|0,f_0\rangle$.}:

\begin{eqnarray}
\langle
\varphi|H_1|\varphi\rangle=\sum_{k'}\varepsilon_{k'}\{\alpha_{k'}^*\alpha_{k'}+
\beta_{k'}^*\beta_{k'}\}\,,\nonumber\\[2ex]
\langle\varphi|H_2|\varphi\rangle= \sum_{k'}\{\beta_{k'}^*\beta_{k'}\}z_R \,,\nonumber\\[2ex]
\langle\varphi|H_3|\varphi\rangle=
\sum_{k'}\alpha_{k'}^*\Lambda_{k'}(\beta)
+\sum_{k'}\beta_{k'}^*\Lambda_{k'}(\alpha)\,, \label{s26}
\end{eqnarray}
where

\begin{eqnarray}
\Lambda_{k'}(\alpha)=\sum_{k''}\Lambda_{k'k''}\alpha_{k''}\,,\hskip0.5cm
\Lambda_{k'}(\beta)=
\sum_{k''}\Lambda_{k'k''}\beta_{k''}\,.\label{s27}
\end{eqnarray}
Using (\ref{s26}) and (\ref{s7}), we finally get:

\begin{eqnarray}
\langle\varphi|H|\varphi\rangle=
\sum_{k'}|\alpha_{k'}|^2\varepsilon_{k'}+
\sum_{k'}|\beta_{k'}|^2(\varepsilon_{k'}+z_R)+\sum_{kk'}(\alpha_{k'}^*
\beta_k+\beta_{k'}^*\alpha_k)\,.\label{s28}
\end{eqnarray}

In order to obtain solutions for the eigenvalue equation
$H\varphi=E\varphi$, we use the following variational principle:

\begin{eqnarray}
\frac{\partial}{\partial\alpha_{k'}^*}\{\langle\varphi|H|\varphi\rangle
-E\langle\varphi|\varphi\rangle\}=0\label{s29}\\[2ex]
\frac{\partial}{\partial\beta_{k'}^*}\{\langle\varphi|H|\varphi\rangle
-E\langle\varphi|\varphi\rangle\}=0\label{s29}\,.
\end{eqnarray}

This gives a  system of linear equations in the undetermined
$\alpha_{k'}$, $\beta_{k'}$, which gives the eigenvector
$|\varphi\rangle$ and $E$. This system is (we are assuming that
$\Lambda_{kk'}=\Lambda_{k'k}$)

\begin{eqnarray}
\alpha_{k'}\varepsilon_{k'}+\sum_{k'}\Lambda_{kk'}\beta_{k'}=E\alpha_{k'}\label{s30}\\[2ex]
\beta_{k'}(\varepsilon_{k'}+z_R)+\sum_{k}\Lambda_{kk'}\alpha_k=E\beta_{k'}\label{s31}\,.
\end{eqnarray}

This can be also written as:

\begin{eqnarray}
\alpha_{k'}(\varepsilon_{k'}-E)=-\sum_k\Lambda_{kk'}\beta_k\,,
\label{s32}\\[2ex] \beta_{k'}(\varepsilon_{k'}+z_R-E)=-
\sum_k\Lambda_{kk'}\alpha_k\,. \label{s33}
\end{eqnarray}
Thus, the eigenvalue equation can be written in the following
matrix form:
\begin{equation}
\left(\begin{array}{cc} {\cal E} & {\bf\Lambda}\\
{\bf\Lambda} & {\cal E}+{\cal Z}
\end{array}\right) \left(\begin{array}{c} {\bf a}\\{\bf b}
\end{array}\right)=E \left(\begin{array}{c} {\bf a}\\{\bf b}
\end{array}\right)\,, \label{s35}
\end{equation}
with

\begin{eqnarray}
{\bf\Lambda}=\left(\begin{array}{cccc}\Lambda_{11} & \Lambda_{12} & \cdots & \Lambda_{1k_n} \\
\Lambda_{21} & \Lambda_{22} & \cdots & \Lambda_{2k_n} \\
\cdots & \cdots & \cdots & \cdots \\
\Lambda_{k_n1} & \Lambda_{k_n2} & \cdots & \Lambda_{k_nk_n}
\end{array}\right) \,, \hskip0.6cm {\cal E}= \left(\begin{array}{cccc}
\varepsilon_{k_1} & 0 & \cdots & 0
\\
0 & \varepsilon_{k_2} & \cdots & 0 \\
\cdots & \cdots & \cdots & \cdots \\
0& 0& \cdots & \varepsilon_{k_n}
\end{array}\right)\label{s36}\\[3ex]
{\cal E}+{\cal Z}= \left(\begin{array}{cccc} \varepsilon_{k_1}+z_R
& 0 & \cdots & 0
\\
0 & \varepsilon_{k_2}+z_R & \cdots & 0 \\
\cdots & \cdots & \cdots & \cdots \\
0& 0& \cdots & \varepsilon_{k_n}+z_R  \end{array}\right)
\label{s37}
\end{eqnarray}

\begin{equation}
{\bf a}=\left(\begin{array}{c} \alpha_1\\ \alpha_2\\ \vdots \\
\alpha_{k_n}
\end{array}\right)\,, \hskip1cm {\bf b}= \left(\begin{array}{c} \beta_1\\ \beta_2\\ \vdots \\
\beta_{k_n}
\end{array}\right)\,,\label{s38}
\end{equation}
where $k_n$ is the total number of fermions considered.

If we have one fermion state, only, we are in the case studied in
the previous subsection. If we have two fermions states, the
eigenvalue equation is a equation of fourth order and therefore
solvable.  In the simplest case in which

\begin{equation}
{\bf\Lambda}=\left(\begin{array}{cc} 0 & a
\\ a & 0 \end{array}\right)\,,
\end{equation}
 $a$ being a complex constant, the eigenvalues of equation (\ref{s35})
fulfill the following equation:

\begin{eqnarray}
\det\left(\begin{array}{cccc}
\varepsilon_1-E & 0 & 0 & a\\
0 & \varepsilon_2-E & a & 0\\
0 & a& \varepsilon_1+z_R-E & 0\\
a & 0 & 0 & \varepsilon_2+z_R-E
\end{array}\right)=0\nonumber\\[2ex]
=\{(\varepsilon_1-E)(\varepsilon_2+z_R-E)-a^2\}\{(\varepsilon_2-E)
(\varepsilon_1+z_R-E)-a^2\}\,.\label{s40}
\end{eqnarray}

The solutions of (\ref{s40}) are

\begin{eqnarray}
E_{1,2}=\frac 12\,
\left(z_R+\varepsilon_1+\varepsilon_2\pm\sqrt{(z_R+\varepsilon_1+\varepsilon_2)^2-
4(\varepsilon_1\varepsilon_2+\varepsilon_2 z_R-a^2)}\right)\\[2ex]
E_{3,4}=\frac 12\,
\left(z_R+\varepsilon_1+\varepsilon_2\pm\sqrt{(z_R+\varepsilon_1+\varepsilon_2)^2-
4(\varepsilon_1\varepsilon_2+\varepsilon_1 z_R-a^2)}\right)\,.
\end{eqnarray}
Obviously, these four solutions are complex.

\subsection{Realistic Applications}

In this subsection we shall rephrase some results of the previous
section in terms of the particle-vibration coupling scheme
\cite{BM2} and, particularly, in the context of the
particle-vibration coupling in Bergreen's basis \cite{cdl}.

To start with, we shall discuss the structure of the
particle-vibration coupling, in a resonant basis and by using
separable two-body interactions. The boson sector of the
Hamiltonian is usually treated in terms of the TDA (Tamm-Damcoff)
approximation, leading to vertex functions of the type \cite{cdl}

\begin{equation}
\Lambda(\omega,\eta)=\sum_\nu
\frac{V^2_{\eta,\nu}}{(e_{\nu}-\omega)}
\end{equation}
where $V$ is the matrix element of the two-body interaction, $\nu$
and $\eta$  denote particle-hole configurations and $\omega$ is
the energy of the one phonon state. If in the configuration $\eta$
(or $\nu$) a single-particle resonant state participates, the
energy of the one phonon state, $\omega$, becomes complex and the
imaginary part of this energy will be proportional to the
imaginary part of the energy of the particle-hole configuration.
Tipically, in a large single particle basis, with few resonant
states, the imaginary part of the energy of a high-lying
one-phonon state is of the order of 1 keV or smaller \cite{cdl}.

In the context of the results of the previous subsection, there is
a direct correspondence between Eq. 51 and $\Lambda(\omega,\eta)$.
Note that, while $\Lambda (z_R)$ has been obtained by working with
the fermion-boson interaction firstly, the result
$\Lambda(\omega,\eta)$ has been obtained (see \cite{cdl}) by
diagonalizing the boson sector and then by using it in the
transformation of the particle-vibration coupling.

The just mentioned analogy suggest that both procedures may be
equivalent. These procedures are:

i.) The treatment of particle-vibration couplings in resonant
(Bergreen) basis. It consists in including fermion resonant states
in single-particle basis, solving for TDA or RPA equations and
getting complex phonon energies, extracting vertex functions and
performing a particle-vibration coupling calculation between
effective fermion and boson fields.

ii.) The extended Friedrichs model: It consists in solving the
Friedrichs model equations proposed in this work, isolating a
boson resonant state, ignoring the background and performing a
particle-vibration coupling scheme by using bound fermions and the
resonant boson.

In the case of the simplified versions of section 3.2, the
solution $E_4$, in (78), closely resembles some of the results of
\cite{cdl}. For the case of one particle above the Fermi level
$E_4$ yields a physical value
\begin{equation}
E_4 \approx \varepsilon_1-\Lambda^2(z_R)
\end{equation}
where we have set $a=\Lambda(z_R)$. Thus, the coupling with the
resonant boson state leads to a imaginary energy of the order of
the imaginary part of the quantity $\Lambda(z_R)$.

\section{Conclusions}

In this work we have presented an extended version of the
Friedrichs model. It consists of couplings between fermion and
bosons, in addition to the couplings between the discrete and
continuous boson excitations. The extended model is exactly
solvable and the structure of the solutions shows a suitable
mechanism for the appearance of complex energies in the fermion
spectrum. In order to show the possible relevance of the present
results for nuclear structure physics we have performed a
comparison with the results obtained in realistic coupling
schemes. The coincidence of the formal structure of the solutions,
for fermion states, in both approaches, suggests a direct
conection between them. The results allow for the interpretation
of the imaginary part of the energy of the fermion states obtained
by coupling fermion bound and resonant states to collective
nuclear particle-hole excitations. Further studies, concerning the
calculation of two-particle correlation functions and
particle-phonon correlation functions in resonant basis are in
progress.

\section{Acknowledgements}

We want to express our gratitude to Prof. G.P. Pronko for his help
in the discussion of the extended Friedrichs model. Financial
support is acknowledged to the Ministry of Education of Spain (PR
2004-0080), the Antorchas foundation (Argentina) and the CONICET
(Argentina).

\end{document}